\documentstyle[12pt]{article}

\newlength{\dinwidth}
\newlength{\dinmargin}
\def\lapproxeq{\lower .7ex\hbox{$\;\stackrel{\textstyle
<}{\sim}\;$}}
\def\gapproxeq{\lower .7ex\hbox{$\;\stackrel{\textstyle
>}{\sim}\;$}}
\def\be{\begin{equation}}
\def\ee{\end{equation}}
\def\bea{\begin{eqnarray}}
\def\eea{\end{eqnarray}}

\makeatletter
\def\fmslash{\@ifnextchar[{\fmsl@sh}{\fmsl@sh[0mu]}}
\def\fmsl@sh[#1]#2{%
\mathchoice
{\@fmsl@sh\displaystyle{#1}{#2}}%
{\@fmsl@sh\textstyle{#1}{#2}}%
{\@fmsl@sh\scriptstyle{#1}{#2}}%
{\@fmsl@sh\scriptscriptstyle{#1}{#2}}}
\def\@fmsl@sh#1#2#3{\m@th\ooalign{$\hfil#1\mkern#2/\hfil$\crcr$#1
#3$}}
\makeatother

\begin{document}
\titlepage
\begin{flushright}
DTP/97/34 \\
hep-ph/9705258 \\
May 1997 \\
\end{flushright}
\begin{center}
\vspace*{2cm}
{\Large \bf Dijet hadroproduction with rapidity gaps and QCD \\
\vspace*{0.1cm}
double logarithmic effects}\\
\vspace*{1cm}
A.D.\ Martin$^a$, M.G.\ Ryskin$^{a,b}$ and V.A.\ Khoze$^{a,c}$

\end{center}
\vspace*{0.5cm}
\begin{tabbing}
$^a$xxxx \= \kill
\indent $^a$ \> Department of Physics, University of Durham,
Durham, 
DH1 3LE, UK. \\

\indent $^b$ \> Petersburg Nuclear Physics Institute, SU-188 350
Gatchina, Russia. \\

\indent $^c$ \> INFN, Laboratori Nazionali di Frascati, PO Box
13, 00044, Frascati, Italy. \\

\end{tabbing}

\vspace*{2cm}
\begin{abstract}
We show that the hadroproduction of a pair of jets with large
transverse energy in the central region bounded by rapidity gaps
is an ideal process to see important double
logarithmic QCD suppression effects.  We compute the cross
sections for both exclusive and inclusive double-diffractive
dijet production at Tevatron and LHC energies for a range of
rapidity configurations.
\end{abstract}
\newpage

\noindent {\large \bf 1.~Introduction} 

Processes with large rapidity gaps in high energy $p p$ (or $p
\overline{p}$) collisions are being intensively studied
both theoretically and experimentally, see for example
refs.\cite{BJ,D0,PUMP,BC}.  One reason is that the
requirement of a rapidity gap is a way to select events induced
by QCD Pomeron exchange.  Another reason is that events with a
rapidity gap offer the opportunity to search for new heavy
particles, such as Higgs bosons, in an environment in which the
large QCD background is suppressed.

A particularly illuminating \lq test' process with
which to probe
the underlying dynamics is the hadroproduction of a dijet system
separated from the beam remnants by rapidity gaps\footnote{Since
we shall assume that the process is mediated by gluon $t$ channel
exchanges our calculation applies equally well to $p
\overline{p}$ and $p p$ collisions.}
\be
\label{eq:a1}
p \overline{p} \rightarrow X \: + \: j j \: + \: \overline{X},
\ee
where the two centrally produced jets each have large transverse
energy $(E_T)$.  The \lq plus' signs in (\ref{eq:a1}) indicate
the existence of
rapidity gaps.  This process has been observed at the Tevatron
\cite{D0}. Moreover, it can be studied in more detail and, in
particular, at larger $E_T$ at the LHC.  Exclusive dijet
production, $p \overline{p}
\rightarrow p +
j j + \overline{p}$, was originally discussed at the Born level
by  Pumplin \cite{PUMP}, and by Berera and Collins \cite{BC}. In
this paper we are concerned
with QCD effects which give significant modifications to the Born
prediction.  Indeed a novel and interesting effect
is the strong suppression of the cross section by double
logarithmic
QCD form factors which reflect the fact that the emission of
relatively soft gluons in the rapidity gap intervals is forbidden
by the experimental cuts.

The first study \cite{KMR} of such a suppression was in
connection with the rapidity gap Higgs signal at the LHC.  It was
found that the cross section for the exclusive process $p p
\rightarrow p + H + p$ was suppressed by Sudakov form factors by
about a factor of 1000 in comparison with the Born cross section.
Here we study the analogous effect in dijet production.  This is
an important process for two reasons.  First, the prediction can
be directly checked by experiment and, second, $b \overline{b}$
dijet production
with rapidity gaps is the main source of QCD background for an
intermediate mass Higgs boson + rapidity gaps signal.  One way in
which
the dijet process
differs from Higgs production \cite{KMR} is that the Sudakov form
factor suppression is partially alleviated by the special
kinematics of the process.  Just as the QED radiative
corrections\footnote{One of the best places to see experimental
evidence of QED double logarithmic effects is in the $J / \psi$
line shape in $e^+e^-$ annihilation.  The asymmetric widening of
the line shape, arising from the radiative tail, is mainly due to
these
effects, see for example \cite{AKLV}.  To obtain a sharp
resonance peak it would be necessary
to experimentally forbid QED radiation from the incoming $e^+$
and $e^-$, which would lead to a Sudakov-suppressed cross
section. 
Clearly the analogous QCD effects which we will discuss here are
much larger.} have to be calculated for each particular choice of
experimental cuts, so the QCD double logarithmic suppression
needs to be evaluated for each specification of the rapidity
gaps.

For pedagogical reasons we first study the exclusive dijet
production process
$p \overline{p} \rightarrow p + j j + \overline{p}$ in which the
proton and antiproton remain intact.  However we find, as
expected, that the cross section is extremely small and so we
then turn to the more realistic inclusive dijet production
process given in (\ref{eq:a1}).  Our study will concentrate on
the
kinematic configuration where the \lq \lq dijet" rapidity
interval between the two rapidity gaps is not large.  Here the
predictions are particularly clear.  If the interval is large
then, as we shall see, the production of additional minijets will
considerably complicate the theoretical framework without giving
any additional insight.  Of course we will also have to take into
account the suppression of rapidity gap events due to
parton-parton rescattering.

We will work in the double logarithmic approximation
and use
the leading power of all logarithms that occur.  Provided that
$E_T$ is sufficiently large, this approach is rather well
justified. \\

\noindent {\bf 2.~Dijet kinematics}
\vspace*{0.3cm} \\
\indent Consider dijet hadroproduction at the quark level, $q q
\rightarrow q + j j + q$, which is shown in Fig.~1.  The
experimental configuration of interest is where particle
production is forbidden in the rapidity gaps
$$
\Delta \eta {\rm (veto)} \; = \; \pm \: (\eta_{{\rm min}},
\eta_{{\rm max}})
$$
and a pair of larger $E_T$ jets are produced with rapidities
$\eta_1, \eta_2$ which both lie in a central rapidity interval
denoted by $\Delta \eta$(dijet).  The rapidity gap configuration
is sketched on the right hand side of Fig.~1.  It is necessary to
choose the interval $\Delta \eta$(dijet) smaller than the
interval $(-\eta_{{\rm min}}, \eta_{{\rm min}})$ between the gaps
so as to ensure that the fragments of the large $E_T$ jets can
be collected and their momenta determined.

Let us denote the transverse momenta of the two jets by
$\mbox{\boldmath$P$}_{1T}$ and $\mbox{\boldmath$P$}_{2T}$.  Then
the jet transverse energies are $E_{iT} =
P_{iT}$.  It is convenient to write the differential cross
section for dijet production in the form $d \sigma / (d^2
\mbox{\boldmath$P$}_T \: d^2 \Delta \mbox{\boldmath$P$}_T \: d
\eta \: d \Delta \eta)$ where
\be
\label{eq:b3}
\mbox{\boldmath$P$}_T \; = \; \textstyle{\frac{1}{2}} \:
(\mbox{\boldmath$P$}_{1T} + \mbox{\boldmath$P$}_{2T}), \quad
\eta \; = \; \textstyle{\frac{1}{2}} (\eta_1 + \eta_2)
\ee
\be
\label{eq:b4}
\Delta \mbox{\boldmath$P$}_T \; = \; \mbox{\boldmath$P$}_{1T} - 
\mbox{\boldmath$P$}_{2T}, \quad \Delta \eta = \eta_1 -
\eta_2.
\ee
On the other hand the $g g
\rightarrow j j$ hard subprocess is most naturally described in
terms of the Mandelstam variables
$\hat{s} = M^2$ and $\hat{t}$, where $M$ is the
invariant mass of the dijet system.  For exclusive dijet
production, $p \overline{p} \rightarrow p + j j + \overline{p}$,
the proton form factors limit the momentum transfer from the
proton (and from the antiproton).  We thus have $|t_i| \ll P_T^2$
and consequently $(\Delta P_T)^2 \ll P_T^2$.  The kinematics are
much simpler when $\Delta P_T$ is small.  It means that the
rapidity axis defined with respect to the direction of the
incoming $p$ and $\overline{p}$ essentially coincides with the
axis defined by the incoming hard gluons.  In this limit
\be
\label{eq:b5}
\hat{s} \: \equiv \: M^2 \: \simeq \: 4 P_T^2 \: {\rm
cosh}^2 (\Delta \eta / 2)
\ee
\be
\label{eq:b6}
\hat{t} \: \simeq \: -P_T^2 (1 + e^{- \Delta \eta})
\ee
where $\Delta \eta$ can be defined by either the $p \overline{p}$
or the $g g$ incoming \lq beams'.

Even for inclusive dijet production, $p \overline{p} \rightarrow
X + j j + \overline{X}$, the condition $(\Delta P_T)^2 \ll P_T^2$
is usually satisfied, unless the experimental criteria insist
otherwise.  Thus we shall assume the above kinematics in this
paper. 

Suppose for the moment we continue to work at the
quark level, $qq \rightarrow q + j j + q$.  We denote the
amplitude for the process by $T$ and write the differential cross
section
\be
\label{eq:c4}
\frac{d \sigma}{d P_T^2 \: d \eta \: d \Delta \eta \: d t_1 \:
dt_2} \; = \; \frac{| T |^2}{(16 \pi^2)^2}.
\ee
Here we have used the identity
\be
\label{eq:c5}
d^2 \Delta P_T \: d^2 q_{1 T}^{\prime} \: d^2 q_{2 T}^{\prime} \;
\delta^{(2)} \: (\Delta \mbox{\boldmath$P$}_T +
\mbox{\boldmath$q$}_{1T} + \mbox{\boldmath$q$}_{2T}) \; = \;
\pi^2 \: d t_1 d t_2
\ee
where $q_1^{\prime}$ and $q_2^{\prime}$ are the momenta of the
outgoing quarks.  The amplitude $T$ is proportional to the
amplitude ${\cal M}$ describing the on-shell $g g \rightarrow j
j$
subprocess.  We normalise ${\cal M}$ by
\be
\label{eq:b2}
\frac{d \hat{\sigma}}{d \hat{t}} \; = \; | {\cal M} |^2.
\ee
To relate $d \hat{t}$ to $d P_T^2$ we note that at fixed $M^2$ we
have
\be
\label{eq:c3}
\frac{d \hat{t}}{d P_T^2} \quad \simeq \quad \frac{1 + e^{-
\Delta
\eta}}{1 -
e^{- \Delta \eta}} \quad \simeq \quad M^2 \: \frac{d (\Delta
\eta)}{d
M^2}.
\ee 
This identity means that the form of the differential cross
section of the subprocess which emerges naturally from
multi-Regge kinematics can be written as
\be
\label{eq:c2}
\frac{d \hat{\sigma}}{d P_T^2} \: \frac{d M^2}{M^2} \; = \;
\frac{d \hat{\sigma}}{d \hat{t}} \: d (\Delta \eta).
\ee

\noindent {\bf 3.~Exclusive dijet production} 
\vspace*{0.3cm} \\
\indent We begin with the calculation of the double diffractive
dijet production where the proton (antiproton) remains intact. 
The Born amplitude for quark-initiated production is described by
the Feynman diagram shown by the solid lines in Fig.~1. We must
explain the origin of the second $t$ channel gluon.  Without the
rapidity gap restriction the dijet system could simply be
produced by gluon-gluon fusion.  However, the colour flow induced
by such a single gluon exchange process would then produce many
secondary
particles which would fill up the rapidity gap.  To screen the
colour flow it is necessary to exchange a second $t$ channel
gluon.  At lowest order in $\alpha_S$ this gluon couples
only to the incoming quark lines.  (The case in which the
screening gluons also couple to the high $E_T$ jets is of higher
order in $\alpha_S$).  Thus the Born amplitude is given by
\be
\label{eq:a2}
T (q q \rightarrow q + j j + q) \; = \; \frac{2}{9} \: 2 \: \int
\:
\frac{d^2 Q_T}{Q^2 k_1^2 \: k_2^2} \: 8 \alpha_S^2 (Q_T^2)
\hat{\cal{M}}
\ee
where $\frac{2}{9}$ is the colour factor for the two-gluon
colour-singlet exchange process and the factor 2 takes into
account that both $t$ channel gluons can radiate the dijet
system, see, for example, \cite{PUMP,KMR}.  The amplitude
$\hat{\cal{M}}$
represents the sum of the Feynman diagrams for the subprocess $g
g \rightarrow j j$.  In addition to the usual Mandelstam
variables $\hat{s}$ and $\hat{t}$, the amplitude $\hat{\cal{M}}$
depends on the transverse momenta $\mbox{\boldmath$k$}_{iT}$ of
the incoming gluons.  We work in the limit where $k_{iT}^2 \ll
E_T^2$.  In this limit the off-shell amplitude is of the form
\be
\label{eq:b1}
\hat{\cal{M}} \; = \; k_{1T} \: k_{2T} \: {\cal M} (\hat{s},
\hat{t}),
\ee
which reflects the gauge invariance of the amplitude.  The
remaining factor ${\cal M}$ is the amplitude which describes the
on-shell $g g
\rightarrow j j$
subprocess, which was introduced in (\ref{eq:b2}) and which fixes
the normalisation of (\ref{eq:a2}) and (\ref{eq:b1}).  In the
leading
log approximation the origin of the $k_{iT}$ factors in
(\ref{eq:b1}) is clear from the well-known
Weizs\"{a}cker-Williams formula.
The QCD analogue can be found in ref. \cite{AP}.  The $k_{iT}$
factors
occur since the forward emission of massless vector particles
without spin flip is forbidden \cite{BFK}.

If the momentum transfers are
small $(k_1^2 \approx k_2^2 \approx Q^2)$ then the integral in
(\ref{eq:a2}) behaves as $\int d Q_T^2/Q_T^4$.  Thus small values
of
$Q_T$ of the screening gluon are favoured.  Fortunately, as we
shall see, the existence of the rapidity gaps and the consequent
Sudakov form factor suppression make the integral infrared
convergent.

The Sudakov form factor $F_S$ is the probability not to emit
bremsstrahlung gluons (one of which is shown by $p_T$ in Fig.~1).
We have
\be
\label{eq:a4}
F_S \; = \; {\rm exp} (-S (Q_T^2, E_T^2))
\ee
where $S$ is the mean multiplicity of bremsstrahlung gluons
\be
\label{eq:a5}
S(Q_T^2,  E_T^2) \; = \; \int_{Q_T^2}^{E_T^2} \: \frac{d
p_T^2}{p_T^2} \: \int_{p_T}^{\frac{1}{2} M} \: \frac{d
\omega}{\omega} \; \frac{3 \alpha_S (p_T^2)}{\pi}
\; = \; \frac{3 \alpha_S}{4 \pi} \: \ln^2
\left(\frac{E_T^2}{Q_T^2} \right).
\ee
Here $\omega$ and $p_T$ are the energy and transverse momentum of
an
emitted gluon in the dijet rest system.  Note that $E_T$ is the
transverse energy of the jet adjacent to the \lq hard' gluon from
which the bremsstrahlung takes place.  The last equality in
(\ref{eq:a5}) assumes a fixed coupling $\alpha_S$ and $E_T \simeq
M / 2$, and is shown
only for illustration.  The lower limit of itegration in
(\ref{eq:a5}) reflects the destructive interference of amplitudes
in which the bremsstrahlung gluon is emitted from a \lq hard'
gluon $k_i$ and from the soft screening gluon $Q$.  That is there
is no emission when the wavelength of the bremsstrahlung gluon
$(\simeq 1/p_T)$ is larger than the separation, $\Delta \rho \sim
1/Q_T$, of the two $t$ channel gluons in the transverse plane,
since then they act as a single coherent colour-singlet system. 
However, the situation is a little more complicated.  As it
stands (\ref{eq:a4}) and (\ref{eq:a5}) represent to double
logarithmic accuracy, the probability to have no bremsstrahlung
at all.  But, in a realistic experiment we do not exclude
bremsstrahlung in the rapidity interval embracing the two larger
$E_T$ jets.  That is in practice it is difficult to distinguish
the bremsstrahlung gluons from gluons which belong to the jets. 
Only emission in some fixed rapidity interval $\Delta \eta$(veto)
is vetoed in an experiment.  For example, the
D0 collaboration \cite{D0} at the Tevatron choose a
rapidity gap interval of 
$\Delta \eta {\rm (veto)} \; = \; (\eta_{{\rm min}} = 2,
\eta_{{\rm max}} = 4.1)$.  The suppression of
(\ref{eq:a5}) should therefore only act in the rapidity interval
$\Delta \eta$(veto).  Now the rapidity of the bremsstrahlung
gluon is $\eta_b \; = \; \ln (\omega /2p_T)$.  The relevant
integration in (\ref{eq:a5}) becomes
\be
\label{eq:a6}
\int \frac{d \omega}{\omega} \: \rightarrow \int d \eta_b
\ee
where we must restrict the $\eta_b$ integration to the rapidity
interval $\Delta \eta$(veto).

In addition to the form factor suppression we must also include
the ladder evolution gluons (shown by the dashed lines in Fig.~1)
and to consider the process at the proton-antiproton level rather
than the quark level.  Both changes are achieved by making the
replacements \cite{RRML}
\be
\label{eq:a7}
\frac{4 \alpha_S (Q^2)}{3 \pi} \; \rightarrow \; f(x, Q^2) \; =
\;
\frac{\partial (xg (x, Q^2))}{\partial \ln Q^2}
\ee
in (\ref{eq:a2}), where $x = x_1$ or $x_2$ for the upper or
lower
ladders in Fig.~1 respectively, and where $f (x, Q^2)$ is the
unintegrated gluon density of the proton.  The identification
(\ref{eq:a7}) is valid for small momentum transfer from the
proton, which is the dominant region for the exclusive process. 
We may therefore set $k_{1T}^2 \approx k_{2T}^2 \approx Q_T^2
\approx Q^2$.  Strictly speaking even at zero transverse
momentum transfer, $q_{1T} - q_{1T}^{\prime} \: = \: 0$, we do
not obtain
the exact gluon structure function, as a non-zero component of
longitudinal momentum is transferred through the two-gluon
ladder. However, in the region of interest, $x \sim 0.01$, the
value of $| t_{{\rm min}} | = m_p^2 x^2$ is so small that we may
safely put $t = 0$ and identify the ladder coupling to the proton
with $f (x, Q^2)$ \cite{RRML}.

When we take the modifications (\ref{eq:a6}) and (\ref{eq:a7})
into account the Born amplitude (\ref{eq:a2}) becomes
\bea
\label{eq:a8}
T (p \overline{p} \rightarrow p + j j + \overline{p}) \; = \;
2 \pi^3 \: \int \: \frac{d Q^2}{Q^4} \: e^{-S (Q_T^2, E_T^2)} \;
f
(x_1, Q^2) f (x_2, Q^2) {\cal M},
\eea 
where $E_T = P_T$.  In the limit $Q_T^2 \ll E_T^2$ the amplitude
${\cal M}$ essentially becomes the on-shell amplitude introduced
in
(\ref{eq:b2}) and (\ref{eq:b1}), which is simply a function of
$\hat{s}$ and $\hat{t}$, and so may be taken outside the loop
integral.  Equation (\ref{eq:a8}) can then be expressed in the
symbolic form
\be
\label{eq:a9}
T \; \equiv \; 16 \pi^2 {\cal L}^{\frac{1}{2}} {\cal M}
\ee
where ${\cal L}$ may be regarded as the pomeron-pomeron
luminosity factor.  Clearly ${\cal L}$ depends on the choice of
the rapidity gap configuration.  The factor $16 \pi^2$ arises due
to the choice of normalisation in (\ref{eq:b2}).

The differential cross section for $p \overline{p} \rightarrow p
+ j j + \overline{p}$ is given in terms of $| T |^2$ by
\be
\label{eq:c1}
\frac{d \sigma}{d P_T^2 \: d \eta \: d \Delta \eta} \; = \;
\frac{| T |^2}{(16 \pi^2)^2 b^2} \; = \; \frac{1}{b^2} \: {\cal
L} \: \frac{d \hat{\sigma}}{d \hat{t}},
\ee
where the last equality follows from (\ref{eq:a9}) and
(\ref{eq:b2}).  To obtain (\ref{eq:c1}) we have integrated over
$d (\Delta P_T)^2$ and the proton momentum transfer, see
(\ref{eq:c5}).  These
integrals are governed by the proton form factors.  We obtain a
factor of $1/b$ from both the proton and antiproton where exp $(-
b | t_i |)$ is taken to be the approximate form of the proton
form factor.

In the standard calculation of the cross section of a hard
scattering process, such as $g g \rightarrow j j$, we would
average over the colours and the polarisations of the incoming
gluons.  Here we have to be more careful.  First, the cross
section $d \hat{\sigma} / d \hat{t}$ describes dijet production
in a colour singlet configuration. Second, in exclusive dijet
production the polarisation vectors of the incoming gluons are
directed along $\mbox{\boldmath$k$}_{iT}$, and
hence are strongly correlated\footnote{The correlation is absent
for inclusive dijet production.} since $\mbox{\boldmath$k$}_{1T}
\simeq \mbox{\boldmath$k$}_{2T} \simeq \mbox{\boldmath$Q$}_T$. 
To determine $d \hat{\sigma} / d \hat{t}$ we perform the
appropriate colour and polarisation averaging and obtain
$$
\frac{d \hat{\sigma}}{d \hat{t}} \; = \; \frac{9}{4} \; \frac{\pi
\: \alpha_S^2 (P_T^2)}{P_T^4},
$$
which is in agreement with the cross section obtained in ref.
\cite{BC}.

It is easy to show that the integral in (\ref{eq:a8}) has a
saddle point given by 
\be
\label{eq:d1}
\ln (E_T^2 / Q^2) \; = \; (2 \pi / 3 \alpha_S (Q^2)) (1 - 2
\gamma)
\ee
where $\gamma $ is the anomalous dimension of the gluon, $f (x,
Q^2) \; \propto \; (Q^2)^{\gamma}$. \\

\noindent{\bf 4.~Inclusive dijet production}
\vspace*{0.2cm} \\
\indent As is usual, the cross section for inclusive production
is expected to be larger than for the exclusive process.  Here
the initial protons may be destroyed and the transverse momentum
$\Delta P_T$ of the dijet system, (\ref{eq:b4}), is no longer
limited by the proton form factor, but it is still smaller than
$P_T$ in the leading log approximation.  The process is shown in
Fig.~2 in the form of the amplitude multiplied by its complex
conjugate.  The partonic quasielastic subprocess is $a b
\rightarrow a^{\prime} + j j + b^{\prime}$.  If the partons $a,
b$ are quarks then the Born amplitude for the subprocess is given
by (\ref{eq:a2}).  However, the form factor suppressions are more
complicated than for the inclusive process.  As the momenta
transferred, $t_i \: = \: (Q - k_i)^2$, are large we can no
longer express the upper and lower \lq blocks' in terms of the
gluon structure function, but instead they are given by BFKL
non-forward amplitudes.

We begin with the expression for the Born cross section for the
subprocess $g g \rightarrow g + j j + g$
\be
\label{eq:d2}
\frac{d \sigma}{d P_T^2 d \eta \: d \Delta \eta} \; = \;
\alpha_S^4 \: \frac{81}{64 \pi^2} \: {\cal I} \: \frac{d
\hat{\sigma}}{d \hat{t}}
\ee
with
\be
\label{eq:d3}
{\cal I} \; = \; \int \frac{d Q^2}{Q^2} \: \frac{d Q^{\prime
2}}{Q^{\prime 2}} \: \frac{d k_{1 T}^2}{k_1^2 \: k_1^{\prime
2}} \: \frac{d k_{2 T}^2}{k_2^2 \: k_2^{\prime 2}} \: k_{1T}
\: k_{1T}^{\prime} \: k_{2T} \: k_{2 T}^{\prime},
\ee
where the six propagators of Fig.~2 are evident.

Again care is needed in the computation of the subprocess cross
section $d \hat{\sigma} / d \hat{t}$.  As we noted from
(\ref{eq:b1}) the off-shell $g g \rightarrow j j$ amplitude is
proportional to the $k_{i T}$ of the incoming
gluons, and the remaining on-shell cross section $d \hat{\sigma}
/ d \hat{t}$ should be therefore computed averaging over gluon
polarisations.  The polarisations are described by vectors
$\mbox{\boldmath$\epsilon$}_i$ which are proportional to the
gluon $\mbox{\boldmath$k$}_{i T}$.  Now the leading log
contribution
comes from the strongly-ordered region $k_{i T} \ll k_{jT}$ with
$i \neq j$.  As before comparatively small values of the momentum
$Q$ of the screening gluon are favoured.  However now, without
the presence of proton form factors, the total momentum transfer
$Q - k_i$ may be large.  In the limit $Q^2 \ll k_i^2$ this means
that $k_i$ has to be balanced by $k_i^{\prime}$ for both $i =
1,2$.  That is we have
\be
\label{eq:d4}
t_i \; = \; (Q - k_i)^2 \; \simeq \; -k_{i T}^2 \; \simeq \; -
k_{i T}^{\prime 2}.
\ee
The consequence for the polarisation averaging is that we require
$\mbox{\boldmath$\epsilon$}_i \: \simeq \:
\mbox{\boldmath$\epsilon$}_i^{\prime}$, but that
$\mbox{\boldmath$\epsilon$}_i$
is no longer correlated to $\mbox{\boldmath$\epsilon$}_j$ (as it
was for
exclusive production).  After averaging, the on-shell $g g
\rightarrow g g$ cross section is found to be 
\be
\label{eq:d5}
\frac{d \hat{\sigma}}{d \hat{t}} \; = \; \frac{\pi \alpha_S^2
(P_T^2)}{P_T^4} \: \frac{9}{2} \: \left(1 - \frac{P_T^2}{M^2}
\right)^2,
\ee
while that for $gg \rightarrow q \overline{q}$ is 
\be
\label{eq:y1}
\frac{d \hat{\sigma}}{d \hat{t}} \; = \; \frac{\pi \alpha_S^2
(P_T^2)}{P_T^2 \: M^2} \; \frac{1}{6} \; \left(1 - \frac{2
P_T^2}{M^2} \right)
\ee
for each flavour of quark.

We have chosen the scale of the coupling $\alpha_S$ to be
$P_T^2$, that is the value which for single inclusive jet
production gave small higher-order corrections and which led to
predictions in
agreement with the data.

Again we must estimate the suppression due to gluon
bremsstrahlung filling up the rapidity gaps.  Now the mean number
of gluons emitted, with transverse momenta $Q_T < p_T < k_{iT}$,
in the rapidity interval $\Delta \eta_i \: = \: \Delta
\eta$(veto) is
\be
n_i \; = \; \frac{3 \alpha_S}{\pi} \: \Delta \eta_i \: \ln
\left ( \frac{k_{iT}^2}{Q_T^2} \right ).
\label{eq:e1a}
\ee
The amplitude for no emission in the gap $\Delta \eta_i$ is
therefore $\exp (-n_i/2)$.  In this way we see that the Born
integral (\ref{eq:d3}) is modified to
\be
{\cal I} \; = \; \int \: \frac{dQ^2}{Q^2} \:
\frac{dQ^{\prime 2}}{Q^{\prime 2}} \: \frac{dt_1}{t_1} \:
\frac{dt_2}{t_2} \: {\rm exp} \biggl(-(n_1 + n_1^{\prime} + n_2 +
n_2^{\prime}
+ S_1 + S_1^{\prime} + S_2 + S_2^{\prime}) / 2 \biggr)
\label{eq:e1}
\ee
where the exponential factor represents the total form factor
suppression in order to maintain the rapidity gaps $\Delta
\eta$(veto).  The Sudakov form factors,
$\exp (- S (k_{i T}^2, E_T^2) / 2) \: \equiv \: {\rm exp} (-S_i /
2)$, arise from the insistence that
there is no gluon emission in the interval $k_{i T} < p_T <
E_T$, see (\ref{eq:a4}) and (\ref{eq:a5}).  Again note that the
Sudakov form factors are multilated due to the imposition of a
specific rapidity gap interval $\Delta \eta$(veto), see the
replacement given in (\ref{eq:a6}).

The justification of the non-Sudakov form factors, $\exp
(-n_i/2)$ is a little subtle.  First we notice from
(\ref{eq:e1}) that due to the asymmetric configuration of the
$t$-channel gluons, $Q_T \ll k_{iT}$, we have, besides $\Delta
\eta_i$, a second logarithm, $\ln (k_{iT}^2 / Q_T^2)$, in the
BFKL
evolution.  These double logs are resummed\footnote{The
resummation corresponds to the Reggeization of the $t$-channel
gluons.} to give the BFKL non-forward amplitude $\exp (-n_i/2)
\Phi (Y_i)$, where the remaining factor $\Phi (Y_i)$ accounts for
the usual longitudinal BFKL logarithms\footnote{Here $\Delta
\eta_i$ (or $Y_i$) plays the role of $\ln (1/x)$ in the BFKL
evolution.},
\be
Y_i \; \equiv \; (3 \alpha_S / 2 \pi) \Delta \eta_i.
\label{eq:e2}
\ee
For rapidity gaps with $\Delta \eta_i \lapproxeq 4$ we have $Y_i
\lapproxeq 0.5$, and it is sufficient to include only the ${\cal
O} (Y_i)$ term, which
gives $\Phi \simeq 1 + Y_i Q_T^2/k_{iT}^2 \simeq 1.1 \pm 0.1$
\cite{FR}.  At our level of accuracy we may neglect the
enhancement due to $\Phi$, and hence we obtain (\ref{eq:e1}),
which is valid in the double $\log$ approximation.

To evaluate ${\cal I}$ of (\ref{eq:e1}) we first perform the
$Q^2$ and $Q^{\prime 2}$ integrations and obtain $(Y_1 + Y_2)^{-
2}$.  Then we integrate over $\ln (t_1/t_2)$ which gives
$\frac{1}{2} (1/Y_1 + 1/Y_2)$ where, at large $\Delta \eta_i$, we
neglect the $t_i$ dependence of $S_i$.  Thus (\ref{eq:e1})
becomes
\be
{\cal I} \; = \; \frac{1}{2 Y_1 Y_2 (Y_1 + Y_2)} \:
\int^{E_T^2} \: \frac{dt}{t} \: \exp (-2 S (t,
E_T^2)).
\label{eq:e3}
\ee
For fixed $\alpha_S$ the final $(dt)$ integration gives $\pi (6
\alpha_S)^{- \frac{1}{2}}$ in the double log approximation. 
However, to predict
the cross section for inclusive production we must
convolute the parton-parton cross sections with the parton
densities $a (x_a, t)$ of the proton, with $a = g$ or $q$, and
evaluate the $dt$ integral numerically.  There is a subtlety when
we come to include these parton luminosity factors
$$
\int_{x_{\rm min}}^1 \: dx_a \: a (x_a, k_{1T}^2) \: \ldots ,
$$
where a summation over $a = g, q$ is implied.  At first sight we
might expect $x_{\rm min} = M / \sqrt{s}$ for central dijet
production.  However,
at large $k_{iT}$ the rapidities of the $a^{\prime}, b^{\prime}$
jets
are small in the dijet rest frame; $\eta_{a^{\prime}} = \ln
(x_{a^{\prime}} \sqrt{s}/k_{1T})$.  Thus in order to maintain the
rapidity gaps $(\eta_{a^\prime} > \eta_{\rm max})$, we must take
\be
(x_1)_{\rm min} \; = \; [M e^{\eta} \: + \: k_{1T} \:
e^{\eta_{{\rm
max}}}] / \sqrt{s},
\label{eq:e4}
\ee \\
and similarly for $(x_2)_{{\rm min}}$.

So far we have considered the simplest $g g \rightarrow j j$ hard
subprocess.  However, if the virtuality $k_i^2$ of the incoming
gluons is much smaller than $P_T^2$ of the outgoing jets, then we
must discuss the
possibility of DGLAP evolution in the $(k_{iT}, P_T)$ interval. 
The evolution means that one (or more) extra jets may be emitted
with transverse momentum $q_T$ in this interval.  At first sight
it appears that the order $\alpha_S$ correction will be enhanced
by a factor $\ln (P_T^2 / k_{iT}^2)$.  Indeed such a contribution
would arise if $\eta_{{\rm min}}$ is sufficiently large so that
$\alpha_S \: \eta_{{\rm min}} \: \sim \: 1$.  The situation can
be described with reference to Fig.~2.  The incoming proton \lq
fragments' into a system of partons with rapidities $\eta >
\eta_{{\rm max}}$.  This process is described by DGLAP evolution
which effectively sums up the collinear logs.  The leading log
contribution comes from the configuration where the angles of the
secondary partons are strongly ordered.  The effect is described
by the parton distribution $a (x_a, k_{1T}^2)$.  The
emission of partons with larger opening angles in the range
$\theta_{{\rm min}} < \theta < \theta_{{\rm max}}$ (corresponding
to the rapidity gap $\eta_{{\rm max}} < \eta < \eta_{{\rm min}})$
is experimentally vetoed.  The resulting suppression of the cross
section is taken into account by the exp$(-S_i / 2)$ and
exp$(-n_i / 2)$ factors in
(\ref{eq:e1}).  Nevertheless,
starting from $\theta \: = \: \theta_{{\rm max}}$, the DGLAP
evolution may be continued to larger angles.  Thus, as well as
the Born process $g
g \rightarrow j j $, we should include more complicated inclusive
subprocesses such as $g g \rightarrow j j g$ shown in Fig.~3. 
Fortunately if $\theta_{{\rm max}}$ is sufficiently large, or
equivalently if $\eta_{{\rm min}}$ is sufficiently small, the
probability of extra jet emission (which is proportional to
$\alpha_S \eta_{{\rm min}}$) may be neglected.  In other words,
the presence of the rapidity gap removes the main part of the
DGLAP enhancement, that is it kills the $\ln (P_T^2 / k_{iT}^2)$
factor which would have occurred due to emission in the rapidity
gap interval in the absence of the experimental veto. \\

\noindent {\large \bf 5.~Predictions for the double-diffractive
dijet cross section}

Our main objective is to estimate the dependence of the cross
section for central dijet production in $p p$ (or $p
\overline{p}$) collisions on the imposition of rapidity gaps.  An
understanding of this double diffractive process is important. 
On the one hand the $b \overline{b}$ dijet channel is  the
background to Higgs production from
either $WW$ or pomeron-pomeron fusion.  On the other hand, dijet
production provides an observable test of novel QCD double
logarithmic
effects.  That is it is possible to study how the event rate
varies according to the experimental choice of the rapidity gaps
in which QCD double logarithmic gluon emission is forbidden. \\

\noindent {\bf 5.1~The inclusive cross section}

Recall that the calculation is done in the leading log
approximation and that we anticipate that the {\it inclusive}
configuration, $p p \rightarrow X + j j + X$, plays the dominant
role.  We present the cross section in the differential form of
(\ref{eq:d2}) for central dijet production with rapidity $\eta \:
\equiv \: \frac{1}{2} \: (\eta_1 + \eta_2) \: = \: 0$.  The
results are shown in Table 1 for both FNAL and LHC energies of
$\sqrt{s} \: = \: 1.8 {\rm TeV}$ and $14 {\rm TeV}$ respectively.

For the FNAL predictions the rapidity gaps were chosen to be
those
used by the D0 collaboration \cite{D0}, that is $\eta_{{\rm min}}
\:
= \: 2$ and $\eta_{{\rm max}} \: = \: 4.1$.  The jets were taken
to have $P_T \: = \: 15 {\rm GeV}$.  At the LHC energy we took
$P_T \: = \: 50 {\rm GeV}$, and besides the above choice of
rapidity gap, we also present results for a larger gap with 
$\eta_{{\rm min}} \: = \: 2, \: \eta_{{\rm max}} \: = \: 6$ so as
to explore the sensitivity to the gap size $\Delta \eta$(veto). 
We calculated the cross section using various recent sets of
partons.  The values in Table 1 were obtained using the MRS(R2)
set of partons \cite{MRS}.

>From the Table we see that, for $\eta \: = \: 0$ and $\Delta \eta
\sim 1$, the inclusive dijet cross sections at LHC and
FNAL are
\bea
\label{eq:f1}
\frac{d \sigma}{d P_T^2 \: d \eta \: d \Delta \eta} \; \approx \;
\left\{ \begin{array}{l l l}
2 \: {\rm pb} / {\rm GeV}^2 & {\rm at} \: \sqrt{s} \: = \: 14
{\rm TeV} &
({\rm with} \: P_T \: = \: 50 {\rm GeV}) \\
100 {\rm pb} / {\rm GeV}^2 & {\rm at} \: \sqrt{s} \: = \: 1.8
{\rm TeV} &
({\rm with} \: P_T \: = \: 15 {\rm GeV}). \\
\end{array} \right.
\eea
The larger value at the Tevatron energy simply reflects the
$1/P_T^4$
behaviour.  Note that the cross sections are rather large.  For
example if we integrate over $P_T^2$ using the above values of
$P_T$ as the lower bounds then the cross sections are
approximately given by multiplying the quoted values by $P_T^2$
in ${\rm GeV}^2$.  That is an integrated cross section of about
5nb at LHC with $P_T > 50 {\rm GeV}$, and 20nb at the Tevatron
with
$P_T > 15 {\rm GeV}$.

The above large cross section values do not take into account the
possibility of multiple parton-parton scattering (see, for
example, ref. \cite{KMR}).  The secondary hadrons produced in
such a rescattering will tend to fill up the original rapidity
gaps.  We thus have to multiply the cross sections in the table
by a factor $W$ which is the probability not to have an inelastic
rescattering.  To estimate $W$ we may use \cite{RR}
\be
\label{eq:f2}
W \; = \; \left(1 - \frac{2 (\sigma_{{\rm el}} + \sigma_{{\rm
SD}} + \sigma_{{\rm DD}})}{\sigma_{{\rm tot}}} \right)^2 \; = \;
0.06 \quad {\rm at} \: \sqrt{s} \: = \: 1.8 {\rm TeV}, 
\ee
where $\sigma_{{\rm el}}, \: \sigma_{{\rm SD}}, \: \sigma_{{\rm
DD}}$ and $\sigma_{{\rm
tot}}$ are the elastic, single and double diffractive and total
$p \overline{p}$ cross
sections respectively.  The numerical estimate in (\ref{eq:f2})
is obtained using the CDF measurements \cite{CDF} of
$\sigma_{{\rm tot}}, \: \sigma_{{\rm el}}$ and $\sigma_{{\rm
SD}}$.  For the double diffractive cross section we use the
factorization relation $\sigma_{{\rm DD}} \; = \; (\sigma_{{\rm
SD}})^2 / \sigma_{{\rm el}}$.  We note that earlier $p
\overline{p}$ cross section measurements \cite{E710} would give
$W = 0.025$.  The smaller value is mainly due to the smaller
measured $\sigma_{{\rm tot}}$.

Alternative ways to estimate $W$ can be
found in refs.\cite{W}.  $W$ is a common overall suppression
factor which affects the cross section for any process with one
or more rapidity gaps\footnote{If for very high energy $p p$
collisions the suppression factor $W$ becomes extremely small,
then the subprocess $\gamma \gamma \rightarrow j j$ could become
competitive.  The reason is that this subprocess arises from
large impact parameters where rescattering is essentially
absent.}.  Thus, although it gives an added
uncertainty to the overall normalisation of the cross section, it
should not modify the form of the $\eta, \: \Delta \eta$ and
$P_T$
dependence.

>From the Table we also see that the $p p \rightarrow X + b
\overline{b} + X$ cross section is about 1\% of the whole cross
section for inclusive double-diffractive dijet production, $p p
\rightarrow X + j j + X$.  That is the dijets are dominantly
gluon-gluon
jets.  As mentioned above, the estimate of $b \overline{b}$
production with rapidity gaps is relevant in assessing the
background to the production of a Higgs boson of intermediate
mass.

The predictions for {\it inclusive} dijet production are stable
to the use of different recent sets of parton distributions and
to different treatments of the infrared region.  The reason is
that the saddle point of the integration of (\ref{eq:e1}) lies in
the perturbative region.  For the LHC energy it occurs at $k_T^2
\simeq 20 {\rm GeV}^2$ and $Q^2 \simeq 4 {\rm GeV}^2$, while for
the FNAL energy it is at $k_T^2 \simeq 2.5 {\rm GeV}^2$ and $Q^2
>
1 {\rm GeV}^2$.  Even in the latter case the uncertainty due to
partons and the infrared contribution is only $\pm 10\%$.

We find that there is about 50\% suppression due to the Sudakov
form factor.  The suppression arising from the presence of the
BFKL non-forward amplitudes is strongly dependent on the size of
the rapidity
gap.  For example at LHC energies the cross section for dijet
production with the larger gap, $\Delta \eta {\rm (veto)} \: = \:
(2, \: 6)$, is a factor of 100 smaller than that for $\Delta \eta
{\rm (veto)} \: = \: (2, \: 4.1)$. 

The rapid decrease with increasing $\Delta \eta$(veto) is a
characteristic feature of perturbative Pomeron effects on this
process.  It comes mainly from the presence of non-forward BFKL
amplitudes which in turn arise from the asymmetric gluon exchange
configurations where Reggeization is important.  It should be
readily observable. \\

\noindent{\bf 5.2~The exclusive cross section}

The calculation of the cross section for the {\it exclusive}
process, $p p \rightarrow p + j j + p$, is much more dependent on
the infrared region.  The problem is that the main contribution
to the integral in (\ref{eq:a8}) comes from rather small values
of $Q$,
even when the Sudakov form factor is included.  The predictions 
are therefore sensitive to the gluon density in the region $Q^2
\approx 1 {\rm GeV}^2$, or less, where it is not well defined. 
Of
course double-diffractive dijet hadroproduction is dominated by
the inclusive process and so the computation of the exclusive
cross section is not so important.  Nevertheless, for
completeness, we give an estimate of the cross section.

The procedure that we follow is similar to that used in
refs.\cite{PUMP,BC}.  The idea is to  use an unintegrated gluon
density $f (x, l_T^2)$ which is truncated at low transverse
momentum $l_T$ in such a way as to reproduce the observed value
of the total (inelastic) $p p$ cross section.  We start from the
Low-Nussinov two-gluon exchange model for the cross section.  The
Born amplitude gives
\be
\label{eq:g1}
\sigma_{pp} \; = \; 4 \pi \alpha_S^2 \: \int \:
\frac{dl_T^2}{l_T^4} \: \frac{2}{9} \: (3) (3),
\ee
where we integrate over the transverse momentum $l_T$ of the
gluons exchanged between 3 (valence) quarks of the protons.  As
usual, $\frac{2}{9}$ is the colour factor.  We may improve this
estimate by rewriting (\ref{eq:g1}) in terms of the unintegrated
gluon density, see the quark level formula (\ref{eq:a7}),
\be
\label{eq:g2}
\sigma_{pp} \; = \; \frac{\pi^3}{2} \: \int \:
\frac{dl_T^2}{l_T^4} \: f (x, l_T^2) \: f (x, l_T^2)
\ee
where $x = 2 l_T / \sqrt{s}$.  In the perturbative region $l_T^2
> l_0^2$ the right hand side is known.  We can therefore insert
the value of the inelastic cross section, $\sigma_{pp} \approx
45$mb, measured at FNAL to determine the infrared contribution to
the integral.  We may either use a \lq \lq sharp cut-off",
putting $f = 0$ for  $l_T^2 < l_0^2$, or employ a \lq \lq soft
cut" by extrapolating into the region $l_T^2 < l_0^2$ using the
linear form
\be
\label{eq:g3}
f (x, l_T^2) \; = \; (l_T^2 / l_0^2) \: f (x, l_0^2).
\ee
To reproduce the observed cross section we find that we need to
take the sharp
cut-off at $l_0 = 1.1 {\rm GeV}$, or alternatively to choose the
soft cut starting at $l_0 = 1.5 {\rm GeV}$.

We assume that this procedure can be taken over to evaluate the
infrared contribution to the exclusive dijet cross section of
(\ref{eq:a8}).  The predicted values of the cross section are
shown in Table 1 for both treatments of the infrared region.  We
see that the exclusive double diffractive dijet cross section is
a factor of about 20 or 150 smaller than the inclusive one at
the Tevatron or LHC energies respectively. The suppression due to
the double logarithmic Sudakov factor is $\frac{1}{2}$ for $P_T =
15 {\rm GeV}$ jets at the Tevatron, whereas it is $\frac{1}{10}$
for $P_T = 50 {\rm GeV}$ jets at the LHC. \\

\noindent{\large \bf 6.~Discussion}

The observation of processes with rapidity gaps is of great
interest for understanding the structure of the perturbative
Pomeron.  In this respect the central production of dijets with a
rapidity gap on either side is an ideal \lq \lq perturbative
laboratory".  The process has a large cross section and may be
studied in
detail at the Tevatron and at the LHC.  Here we have obtained a
formalism
which allows an estimate of the cross section and which
systematically takes into account the main effects, some of which
have not been considered before.  The dijet system is produced by
the fusion of two gluons (of momenta $k_1$ and $k_2$).  A second
$t$ channel gluon exchange (of momentum $Q$) is needed to
neutralise the colour flow.  The main contribution to the cross
section comes when the screening gluon is comparatively soft
$(Q^2 \ll k_i^2)$, yet $Q^2$ is large enough to allow the
inclusive cross section to be reliably estimated by perturbative
QCD.  The cross section is found to be suppressed by Sudakov form
factors and by non-forward BFKL amplitudes, the latter arising
from the asymmetric two-gluon exchange configuration.

We presented sample results for the cross section which
demonstrate the scale of the effects.  We chose the rapidity gaps
to correspond to those used by the D0 collaboration \cite{D0} at
FNAL.  However, at the LHC energy we also presented results for
larger rapidity gaps.  The main uncertainty is from rescattering
effects, which will populate the gaps.  The cross sections
presented in Table 1 do not include the suppression factor
arising from the requirement to have no rescattering.  We gave an
estimate of this factor in (\ref{eq:f2}).

Finally we emphasize that we have worked at the partonic level. 
That is the rapidity intervals are defined with respect to the
emitted partons.  In a realistic experimental situation our
results therefore correspond to smaller rapidity gaps for the
hadrons since a gluon produced just outside the $\Delta
\eta$(veto) interval may produce a secondary inside the gap.  To
obtain an indication
of the size of the effect we recomputed the cross section at the
Tevatron energy (for $\eta = 0, \: \Delta \eta = 0, \: P_T = 15
{\rm
GeV}$) with $\eta_{{\rm min}} \rightarrow \eta_{{\rm min}} - 0.5$
and $\eta_{{\rm max}} \rightarrow \eta_{{\rm max}} + 0.5$ and
found that it was suppressed by a further factor of 12. 

The rescattering and rapidity gap broadening effects give the
main uncertainties in our cross section estimates.  The first
estimates indicate that the combined effect will diminish the
perturbative predictions in Table 1 by more than two orders of
magnitudes.  The broadening of the gap can be simulated by Monte
Carlo studies and can be readily accounted for when the data is
analysed.  However the rescattering estimate is model dependent
and requires confirmation by studying multiplicity distributions
and the energy behaviour of diffractive cross sections. \\

\noindent {\large \bf Acknowledgements}

\noindent We thank Bob Hirosky and Dino Goulianos for discussions
concerning the data, and M.\ Heyssler and M.R.\ Whalley for
useful information.  VAK
thanks PPARC, and MGR
thanks the Royal Society, INTAS (95-311) and the Russian
Fund of Fundamental Research (96 02 17994), for support. \\

\newpage
\begin{table}
\begin{center}
\begin{tabular}{|c||c|c|c|c||c|} \hline
\rule[-0.1cm]{0mm}{1cm}$\Delta \eta$ & $\sigma _{{\rm inc}}$ &
$\sum$ \underline{$\sigma_{{\rm inc}}^{q
\overline{q}}$}
& $\underline{\sigma_{{\rm inc}}^{b \overline{b}}}$ &
$(k_T^2)_{{\rm sad.pt}}$ &
$\sigma_{{\rm exc}}$ \\
& $({\rm pb} / {\rm GeV}^2)$ &
\raisebox{1.2ex}[0pt]{{\scriptsize
$q$}}
$\sigma_{{\rm inc}}$ &
$\sigma_{{\rm inc}}$ & $({\rm GeV}^2)$ & $({\rm pb} / {\rm
GeV}^2)$ \\ \hline
\multicolumn{2}{|c|}{\rule[-0.4cm]{0mm}{1cm}LHC $(\sqrt{s} = 14
{\rm TeV}):$} & & & & \\
0 & 2.4 (0.021) & 4\% & 0.8\% & 20 & 0.026 (0.027) \\
1 & 2.4 (0.019) & 3\% & 0.7\% & 22 & 0.020 (0.021) \\
2 & 2.0 (0.014) & 2\% & 0.4\% & 28 & 0.010 (0.011) \\ \hline
\multicolumn{2}{|c|}{\rule[-0.4cm]{0cm}{1cm}FNAL $(\sqrt{s} = 1.8
{\rm TeV}):$} & & & & \\
0 & 110 (8.7) & 4\% & 0.8\% & 2.5 & 3.9 (5.5) \\
1 & 110 (8.2) & 3\% & 0.6\% & 2.7 & 2.9 (4.2) \\
2 & 88 (5.8) & 2\% & 0.3\% & 3.4 & 1.2 (1.9) \\ \hline
\end{tabular}
\end{center}
\end{table}
\noindent Table 1:  The double-diffractive inclusive dijet cross
section $\sigma_{{\rm inc}}$ as
in (\ref{eq:f1}) evaluated at $\eta = 0$ for three values of
rapidity difference of the jets $\Delta \eta \; = \; \eta_1 -
\eta_2$. The first number for $\sigma_{{\rm inc}}$ corresponds to
the rapidity gap choice
$\Delta \eta ({\rm veto}) = (2, 4.1)$, whereas the number in
brackets corresponds to $\Delta \eta
({\rm veto}) = (2,6)$ for LHC and (1.5, 4.6) for the Tevatron. 
Also shown are the percentage of events where the dijets are
quark-antiquark pairs summed over all types of quarks, and the
percentage of $b
\overline{b}$ events.  $(k_T^2)_{{\rm sad.pt}}$ is the saddle
point of the integration of (\ref{eq:e1}).  The final column is
the exclusive dijet cross section, first evaluated using a sharp
cut-off in (\ref{eq:a8}) and, second (in brackets), using the
soft cut-off of (\ref{eq:g3}).  The transverse momenta
of the
jets are taken to be $P_T = 50 (15) {\rm GeV}$ at LHC(Tevatron)
energies.
\newpage
\noindent {\large \bf Figure Captions}
\begin{itemize}
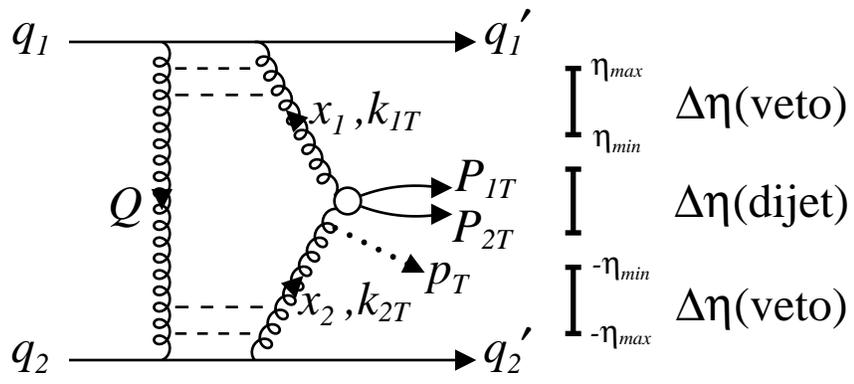
\item[Fig.~1] The Born amplitude for exclusive double-diffractive
dijet production shown at the quark level, together with the QCD
radiative corrections arising from \lq evolution' gluons (dashed
lines) and the Sudakov-type form factor suppression associated
with the $\Delta \eta$(veto) rapidity gaps.  The comparatively
soft screening gluon has four-momentum $Q$.  The rapidity gaps
are indicated by $\Delta \eta$(veto) and the two large $P_T$ jets
are required to lie in the rapidity interval $\Delta
\eta$(dijet).

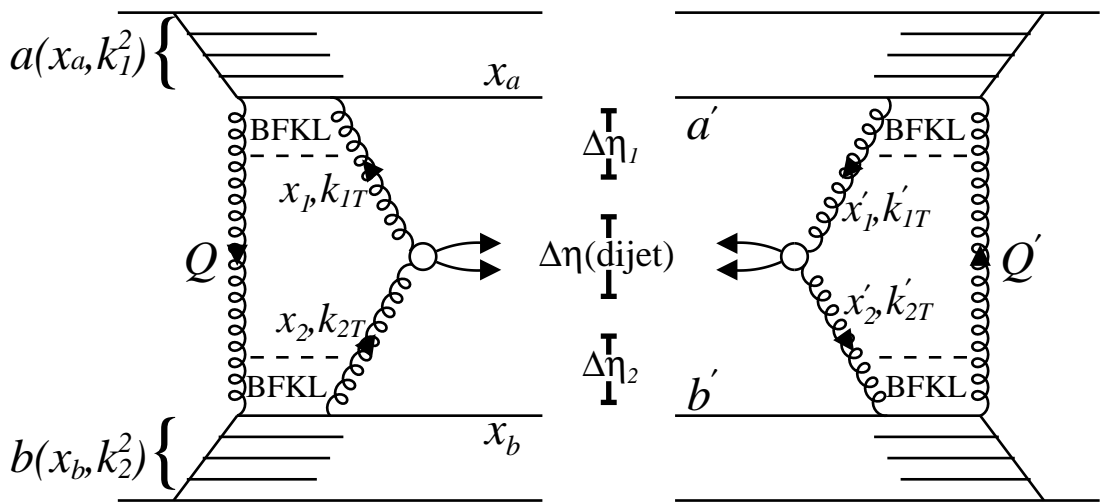
\item[Fig.~2] The amplitude multiplied by its complex conjugate
for inclusive double-diffractive dijet production with rapidity
gaps $\Delta \eta_1$ and $\Delta \eta_2$ on either side, $\Delta
\eta_i \: = \: \pm \; (\eta_{{\rm min}}, \eta_{{\rm max}})$. The
suppression due to QCD radiative effects comes from the double
log resummations exp$(-n_i / 2)$ in the BFKL non-forward
amplitudes and from the Sudakov-like form factor suppressions
exp$(-S_i / 2)$ associated with the rapidity gaps along the hard
gluon lines.
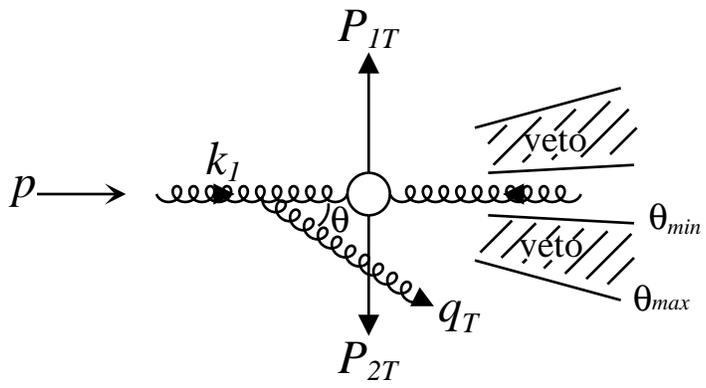
\item[Fig.~3] A schematic diagram of the subprocess $g g
\rightarrow j j g$, where the additional jet with transverse
momentum $q_T$ is emitted with $\theta > \theta_{{\rm max}}$. 
Only the vetoed region in the forward direction is shown.  There
is also a rapidity gap in the backward hemisphere.
\end{itemize}
\end{document}